\documentclass[aps,prd,preprintnumbers,nofootinbib,twocolumn]{revtex4}
\usepackage{bm}
\usepackage{latexsym}
\usepackage{dcolumn}
\usepackage{amsmath,amsfonts,amssymb}
\usepackage{graphicx,epsfig}
\usepackage{color}
\usepackage{amsthm}

\def\be {\begin{equation}}
\def\ee {\end{equation}}
\def\bea {\begin{eqnarray}}
\def\eea {\end{eqnarray}}
\def\bc {\begin{center}}
\def\ec {\end{center}}
\def\bfg {\begin{figure}}
\def\efg {\end{figure}}
\def\bi {\begin{itemize}}
\def\ei {\end{itemize}}
\def\nn {\nonumber}

\def\la {\label}
\def\le {\left}
\def\ri {\right}

\def\D  {\Delta}

\def\m  {\mu}

\def\t  {\tau}

\def\beq{\begin{equation}}
\def\eeq{\end{equation}}
\def\br{\begin{eqnarray}}
\def\er{\end{eqnarray}}
\newcommand{\eel}[1] {\label{#1}\end{equation}}

\newcommand{\bdm}{\begin{displaymath}}
\newcommand{\edm}{\end{displaymath}}

\begin{document}
\title{Black Hole Remnant from Gravity's Rainbow }

\author{Ahmed Farag Ali} \email[]{ahmed.ali@fsc.bu.edu.eg;afarag@zewailcity.edu.eg}

\affiliation{Center for Fundamental Physics, Zewail City of Science and Technology, Giza, 12588, Egypt.\\}
\affiliation{Dept. of Physics, Faculty of Sciences, Benha University, Benha, 13518, Egypt.\\}

\begin{abstract}

In this work, we investigate black hole (BH) physics in the context of gravity rainbow.
We investigate this through rainbow functions that have been proposed by
Amelino-Camelia, et el. in \cite{amerev, AmelinoCamelia:1996pj}.
This modification will give corrections to both the
temperature and the entropy of BH and hence it changes the
picture of Hawking radiation greatly when the size of BH approaches the Planck
scale. It prevents BH from total evaporation,
predicting the existence of BH remnant which may resolve the catastrophic behavior
of Hawking radiation as the BH mass approaches zero.
\par\noindent

\end{abstract}

\maketitle

\section{Introduction}

One common feature among most of  semi-classical approaches to quantum gravity
is the departure from the relativistic dispersion relation by redefining the
physical momentum or physical energy at the Planck scale, or Lorentz invariance
violation. The source of this departure comes from many approaches
such as spacetime discreteness \cite{'tHooft:1996uc}, spontaneous symmetry
breaking of Lorentz invariance in string field theory\cite{LIstring}, spacetime foam models \cite{AmelinoCamelia:1997gz} or
spin-network in Loop quantum gravity (LQG) \cite{Gambini:1998it}. Besides,
there are other approaches such as non-commutative geometry \cite{Carroll:2001ws}
which predicts a Lorentz invariance violation.
All these indications suggest that Lorentz violation may be an essential property in constructing
quantum theory of gravity. Theoretically, the departure from Lorentz invariance
is expressed in a form of modified dispersion relations (MDR). MDR could be an indication for threshold anomalies that might occur in ultra-high energy cosmic rays and TeV photons \cite{AmelinoCamelia:1997gz,AmelinoCamelia:1997jx,AmelinoCamelia:1999wk}. Modern observatories recently are gaining the sensitivity needed to measure these effects, and should be improved in coming few years\footnote{ Threshold anomalies are only predicted by MDR scenarios with preferred reference frame in which they show a full violation of relativistic symmetries and in the same they are not predicted by scenarios in which MDR appears due
to deformation of relativistic symmetry with no preferred reference frame\cite{AmelinoCamelia:2002dx}.}.
For a recent detailed review along the mentioned lines can be found in \cite{amerev}.\par
One of the most interesting class of MDRs, has been suggested in \cite{amerev, AmelinoCamelia:1996pj}  which in the high-energy regime takes the form:

\bea
m^2\simeq E^2-\vec{p}^2+ \eta \vec{p}^2 \le(\frac{E}{E_{p}}\ri)^n \label{MDR}
\eea
where $E_{p}$ describes the energy scale at which the dispersion relation is modified and it
is taken to be the Planck energy, while $n$ and $\eta$ represents a free parameters
characterizing the deviation from Lorentz invariance and $n$ indicates how much strongly the
magnitude of the deformation that is suppressed by $E_{p}$. This formula is compatible with some of the results
obtained in the Loop-Quantum-Gravity approach and reflects some results obtained for theories in
$\kappa$-Minkowski noncommutative spacetime \cite{amerev}. For discussion about
phenomenological implications of Eq. (\ref{MDR}), it is very useful to consult the discussion
after Eq. (13) in the detailed review \cite{amerev}.


A theory that predict naturally MDR is known as double special relativity (DSR)\cite{AmelinoCamelia:2000mn}.
DSR is considered as an extension for special theory of relativity and it extends the invariant quantities to be the Planck energy scale beside the speed of light. The simplest realizations of the idea of DSR
are based on a non-linear Lorentz transformation in momentum space, which imply a deformed Lorentz symmetry such that
the usual dispersion relations in special relativity may be modified by Planck scale corrections.
It should be mentioned that Lorentz invariance violation and Lorentz invariance
deformation are in general conceptually different scenarios. Here we are going to adopt Lorentz invariance
deformation scenarios by considering DSR and its extension in models of rainbow gravity.

In the framework of DSR the definition of
the dual position space suffers a nonlinearity
of the Lorentz transformation. To resolve this issue,
Magueijo and Smolin \cite{Magueijo:2002xx} proposed a doubly general relativity which
assumes that the spacetime background felt by a test particle would depend on its energy. Therefore, there will not be a single metric describing spacetime, but a one parameter family of metrics which depends on the energy (momentum) of these test particles, forming
a {\it rainbow} of metrics (i.e rainbow geometry).
This approach is known as \emph{Gravity Rainbow} and can be mathematically constructed as follows;
the non-linear of Lorentz transformation leads to the following modified dispersion relation
\be
E^2 f(E/E_{p})^2- p^2 g(E/E_{p})^2= m^2 \label{RainbowDisper}
\ee

where $E_{p}$ is the Planck energy scale, $m$ is the mass of the test particle,
 $f(E/E_{p})$ and $g(E/E_{p})$ are commonly known as Rainbow functions and they satisfy $\lim_{E\rightarrow 0} f(E/E_P) =1$
and $\lim_{E\rightarrow 0} g(E/E_P) =1$.\par
A modified equivalence principle was proposed in \cite{Magueijo:2002xx} which requires that one parameter family of energy dependent orthonormal frame fields describe a one parameter family of energy dependent metrics given by
\begin{equation}
h(E/E_P) = \eta^{ab} e_a(E/E_{p}) \otimes e_b(E/E_{p})
\end{equation}
where $e_0(E/E_{p}) =(1/f(E/E_{p})) \tilde{e}_0$ and $e_i(E/E_{p})= (1/f(E/E_{p})) \tilde{e}_i$. But in the limit $(E/E_{p})\rightarrow 0$ general relativity must be recovered. With the definition of one parameter
family of energy momentum tensors Einstein's equations are also modified as
\begin{equation}
G_{\mu \nu}(E/E_{p}) = 8\pi G T_{\mu \nu}(E/E_{p}) \label{RFE}
\end{equation}
Potential investigations on the gravity rainbow can be found in \cite{Galan:2004st}.\par
The choice of the Rainbow functions $f(E/E_{p})$ and $g(E/E_{p})$ is important for making predictions. Among different arbitrary choices in \cite{Galan:2004st,FRWRainbow}, many aspects of the theory have been studied with Schwarzschild metric, black holes, FRW universe, and cosmological
eras such as inflation and scale invariant fluctuations \cite{Barrow:2013gia}. Besides,
we studied recently the possibility of resolving big bang singularity using gravity rainbow in \cite{Awad:2013nxa}.
 In this letter we  continue our investigation about the effect
 of gravity rainbow on BH thermodynamics. We employ the modified dispersion relation of Eq.(\ref{MDR}) \cite{amerev, AmelinoCamelia:1996pj}, which fix the rainbow functions
$f(E/E_{p})$ and $g(E/E_{p})$ and use these rainbow functions to study
the BH thermodynamics and investigate it new properties.
We find that end-point of Hawking radiation is not catastrophic anymore in rainbow BH.
We found that an existence of BH remnants at which the specific heat
vanishes and, therefore, the BH cannot exchange heat with the
surrounding space.

An outline of this paper is as follows. In section (\ref{standard}), We review
briefly how standard Hawking temperature can be obtained from
the Schwarzschild metric and show briefly the \textbf{\emph{catastrophic evaporation}} of BH.
In section (\ref{rainbow}), we investigate BH
thermodynamics in gravity rainbow. We give our conclusions in section (\ref{conclusions}).
\section{Standard hawking radiation}
\label{standard}

Let us first review briefly the standard Hawking radiation process.
Since Hawking temperature is defined in terms of surface gravity $\kappa$ \cite{Hawking:1974sw} as follows:

\bea
T_H= \frac{\kappa}{2\pi} \label{HawkingT}
\eea
where the surface gravity $\kappa$ is defined as \cite{GRbooks}

\bea
\kappa= \lim_{r\rightarrow R_S} \sqrt{-\frac{1}{4} g^{rr}g^{tt} \le(g_{tt},r\ri)^2} \la{surface}
\eea
where $R_S= 2 G M$ is the Schwarzschild radius. In general relativity the surface gravity
is calculated from Eq. (\ref{surface}) for Schwarzschild metric and it is given by
\bea
\kappa= \frac{1}{4MG}
\eea
and hence the Hawking temperature is given by

\bea
T_H= \frac{1}{8 \pi G M}\label{hawT}
\eea
The BH entropy can be calculated through the first law
of BH thermodynamics:

\be dM= T d S \, . \label{TD1} \ee
By integrating Eq.\ (\ref{TD1}) using Eq.\ (\ref{hawT}), one
can obtain the the Bekenstein entropy\cite{Bekenstein:1973ur}
as follows:
\be
S=4~\pi~G~M^2.\la{entropy}
\ee
The specific heat can be calculated using the thermodynamical
relation

\be {\cal C}= T \frac{\partial S}{\partial T}= T  \frac{\partial S}{\partial M} \frac{\partial M}{\partial T}=
\frac{\partial M}{\partial T}\,,\la{SPH}
\ee
By differentiating Eq.\ (\ref{hawT}) and substituting this into
Eq.\ (\ref{SPH})  , the specific heat could be given by

\be
{\cal C} = -8 \pi G
 M^2\,, \la{C0} \ee

The Hawking temperature $T_H$ can be used in the calculation of
the emission rate. The emission rate might be calculated using
Stefan-Botlzmann law considering the energy loss was dominated by
photons.
The emission rate of the BH will be:
\be
\frac{dM}{dt}=-M_{p}^3\frac{\m^{\prime}}{t_{p}}~M^{-2}\,, \la{rate1}
\ee
where $t_{p}=G^{1/2}$ is
the Planck time in natural units, and the form of $\mu$ can be found in
\cite{Adler,Cavaglia:2003qk}. The
exact calculation should consider the squeezing of the fundamental
cell in momentum space, which  modify the emission rate equation (\ref{rate1}). However,
one can neglect this effect in the first order approximation \cite{Niemeyer:2001xk}.
The decay time of the BH can be obtained by integrating
Eq.\ (\ref{rate1}) to give

\be
\t = \le(\frac{1}{3} \ri) \frac{t_{p}}{M_{p}^3} \m^{\prime -1}M^{3}~\,,
\la{decayt}
\ee
One notice that the calculated Hawking temperature $T_H$, Bekenstein
entropy $S$, specific heat ${\cal C}$, emission rate
$\frac{dm}{dt}$, and decay time $\t$ lead to
\emph{\textbf{catastrophic evaporation} }as $m \rightarrow 0$.
This can be explained as following. Since ${\cal C}=0$ only
when $m=0$, the BH will continue to radiate until
$m=0$. But as the BH approaches zero mass, its
temperature approaches infinity with infinite radiation rate.
This was just a brief summary for the Hawking
radiation and the
catastrophic implications of Hawking radiation as the BH mass approaches zero. In the next  section, we study
BH thermodynamics if gravity rainbow is taken into consideration.


\section{Rainbow black hole}
\label{rainbow}

\begin{figure}
\includegraphics[scale=0.35,angle=0]{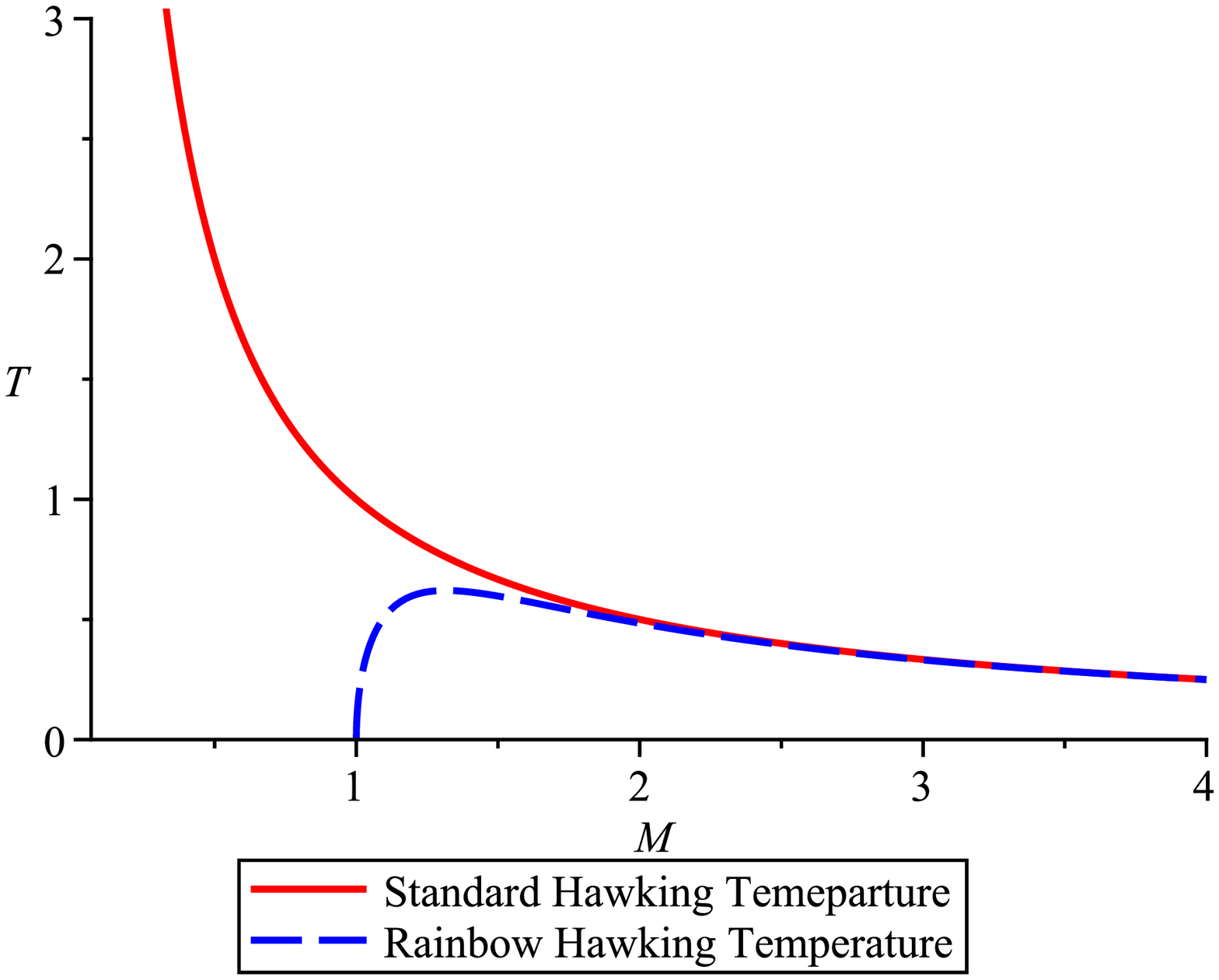}
\caption{$T ~~\text{and}~~ T^{\prime}$ versus $M$}.
\label{Temperature}
\end{figure}

Now we employe the modified dispersion relation which is proposed by the Amelino-Camelia, et al. in Eq. (\ref{MDR}),
and compare it with Eq. (\ref{RainbowDisper}). The functions $f(E/E_p)$
and $g(E/E_p)$ can be fixed as follows:

\bea
f(E/E_p)=1,~~~~~~~~g(E/E_p)=\sqrt{1-\eta \le(\frac{E}{E_{p}}\ri)^n}. \label{RainFunc}
\eea
Let us consider the Rainbow Schwarzschild metric for non-rotating and uncharged  BH \cite{Magueijo:2002xx}

\bea
ds^2&=& -\frac{1}{f\le(E/E_{p}\ri)^2} \le(1-\frac{2MG}{r}\ri) dt^2 \nn\\&&+ \frac{1}{g(E/E_{p})^2}\le(1-\frac{2MG}{r}\ri)^{-1} dr^2
\nn\\&&+\frac{r^2}{g(E/E_{p})^2} d\Omega^2 \label{metric}
\eea
By using the rainbow Schwarzschild metric of Eq. (\ref{metric}), the surface gravity of Eq. (\ref{surface}) get the following form:

\bea
\kappa^{\prime}= \frac{g(E/E_{p})}{f(E/E_{p})} \frac{1}{4 MG} \label{Rainsur}
\eea
If we set $f(E/E_{p})=g(E/E_{p})=1$, then we will get back the surface gravity for non-rotating and uncharged BH.
We find that Eq. (\ref{Rainsur}) introduces the rainbow surface gravity. Then, we can calculate
the modified Hawking Temperature in rainbow gravity  using the modified surface gravity of Eq. (\ref{Rainsur})   as follows:

\bea
T_H^{\prime}= \frac{\kappa^{\prime}}{2\pi} = \frac{g(E/E_{p})}{f(E/E_{p})} \frac{1}{8 \pi MG}
\eea
Using the identification of Eq. (\ref{RainFunc}), the modified Hawking temperature will be given as follows:

\bea
T_H^{\prime}= \sqrt{1-\eta \le(\frac{E}{E_{p}}\ri)^n}\frac{1}{8 \pi MG}\label{modH}
\eea
Let us consider the ordinary uncertainty relation to photons near the event horizon.
According to \cite{Eliasentropy0,Eliasentropy1,Adler,Cavaglia:2003qk} a photon is used to ascertain the position of a quantum particle of energy $E$
and according to the argument in \cite{AmelinoCamelia:2004xx0} which states that the uncertainty principle
$\D p \geq 1/ \D x$  can be translated to the lower bound $E\geq 1/ \D x$
\bea
E\geq \frac{1}{\D x}\simeq   \frac{1}{2 G M} \label{energy}
\eea
where the value of $\Delta x$ has its minimum value taken to be Schwarzschild radius $R_S$, where this is
probably the most sensible choice of length scale in the
context of near-horizon geometry \cite{Eliasentropy0,Eliasentropy1,Adler,Cavaglia:2003qk}.
By substituting Eq. (\ref{energy}) into the modified Hawking Temperature in Eq. (\ref{modH}), we get:
\bea
T_H^{\prime}= \frac{1}{4\pi \le(2 G M\ri)^{\frac{n+2}{2}}}\sqrt{(2 G M)^n- \frac{\eta}{E_p^n}} \label{modT}
\eea
It is clear from Eq. (\ref{modT}) that if we set $\eta=0$ or assume that $E/E_p\rightarrow 0$ (i.e $E_p\rightarrow \infty$), we get
back the Hawking temperature of Eq. (\ref{hawT}). It is also clear from  Eq. (\ref{modT})
that the modified Hawking temperature is physical as far as the BH mass satisfies
the following inequality:

\bea
M\geq \frac{1}{2G} \frac{\eta^{1/n}}{E_p} = \frac{1}{2}\eta^{1/n} E_p
\eea
where we have used the natural units which says $G = 1/ E_p^2$.
This implies that  the black hole should have minimum mass $M_{min} $ given by
\bea
M_{min}=\frac{1}{2}\eta^{1/n} E_p
\eea
This point to an existence of \emph{black hole remnant} due to gravity rainbow.
We plot in Fig.(\ref{Temperature}) the relation between The temperature and BH mass for
both cases of standard and rainbow BH. It is clear that $T$ does not blow up as $M\rightarrow 0$
in rainbow BH in the contrary to the standard BH case. In this figure, we set $n=4$ as an example.
However, different values of $n$ will give similar behavior like in Fig.(\ref{Temperature}) .

The entropy of rainbow BH can be calculated through the first law
of BH thermodynamics:

\be d S= \frac{dM}{T}=\frac{dM}{ \frac{1}{4\pi \le(2 G M\ri)^{\frac{n+2}{2}}}\sqrt{(2 G M)^n- \frac{\eta}{E_p^n}}}  \, . \label{TD2} \ee
With putting $n=4$, and taking $G= 1/E_p^2$ in natural units, the exact form of the entropy will be

\bea
S^{\prime}=\pi {\frac {\sqrt {16\,{M}^{4}-\eta\,{E_{{p}}}^{4}} }{{E_{{p}}}^{2}}}
\eea

\begin{figure}
\includegraphics[scale=0.35,angle=0]{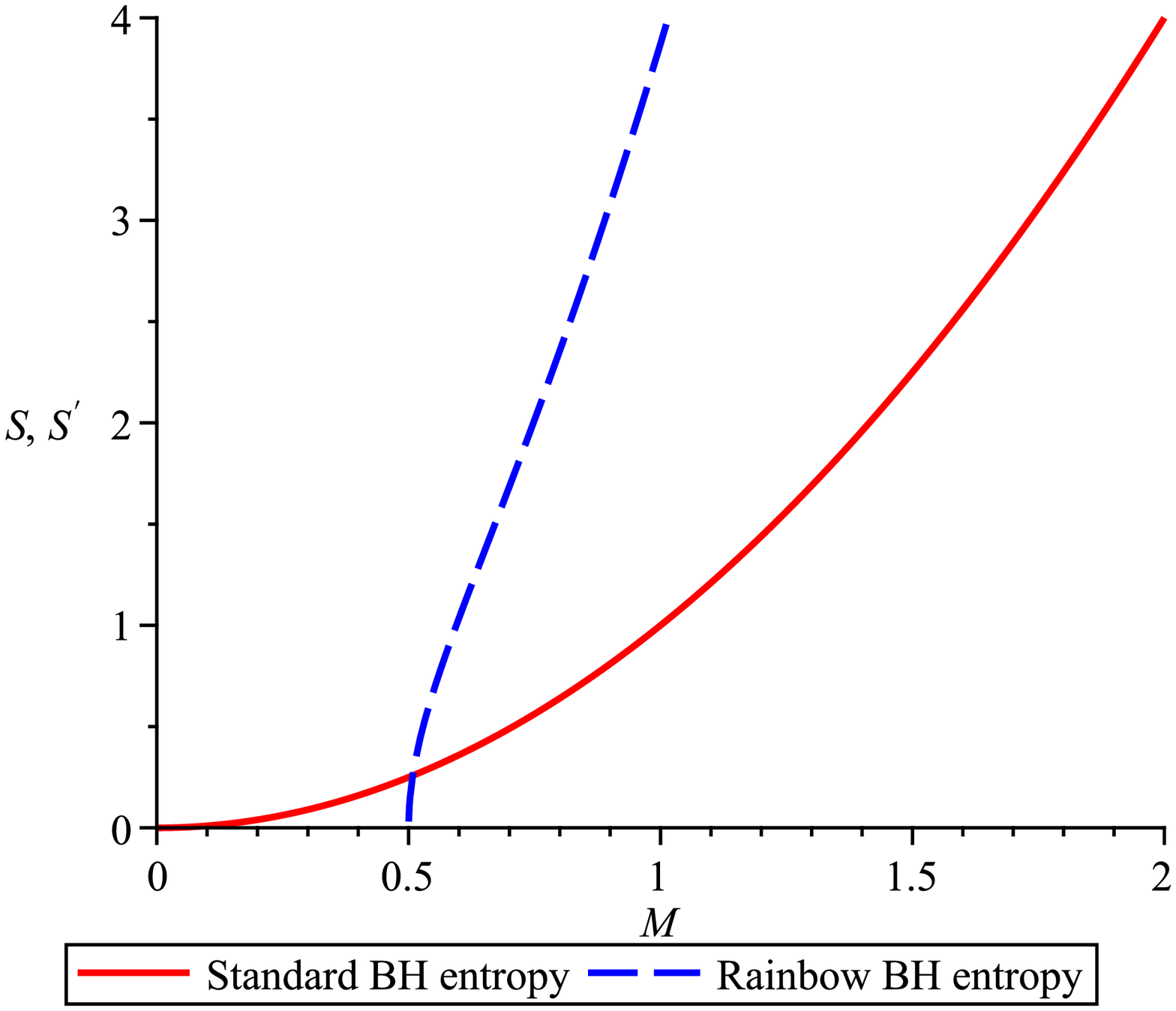}
\caption{$S~~ \text{and} ~~S^{\prime}$ versus $M$}.
\label{Entropy}
\end{figure}

We show in Fig. (\ref{Entropy}), that the entropy of rainbow BH reaches a minimal value which
represents the information contained in the BH remnant.

Turning to calculate the specific heat of the rainbow BH. It is given by

\bea
\mathcal{C}^{\prime}&=& T\frac{\partial S}{\partial T}= \frac{\partial M}{\partial T}\nn\\
&=& -\frac{16\,\sqrt {{2}^{n} \left( {\frac {M}{{E_{{p}}}^{2}}} \right) ^{n}-\eta
\,{E_{{p}}}^{-n}}\pi \,{M}^{2}{2}^{\frac{1}{2\,n}} \left( {\frac {M}{{E_{{p}}}
^{2}}} \right) ^{\frac{1}{2\,n}}
}{{E_{{p}}}^{2} \left( -n\eta\,{E_{{p}}}^{-n}+{2}^{1+n} \left( {\frac {M
}{{E_{{p}}}^{2}}} \right) ^{n}-2\,\eta\,{E_{{p}}}^{-n} \right)
}\nn\\ \label{hawC}
\eea
The last expression for the specific heat of Eq. (\ref{hawC}) indicates that
the specific heat vanishes at $M=M_{min}=\frac{1}{2}\eta^{1/n} E_p$
Therefore, the BH cannot exchange heat with the surrounding space and hence predicting the existence of
\emph{black hole remnants}.

\begin{figure}
\includegraphics[scale=0.35,angle=0]{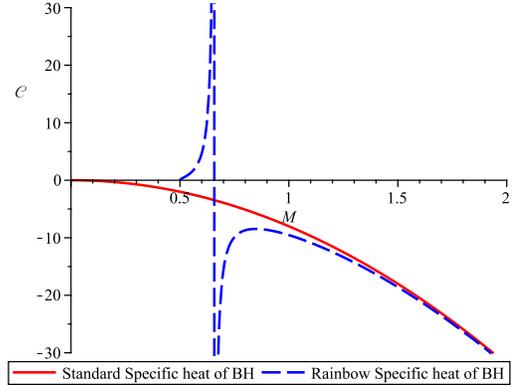}
\caption{$\mathcal{C}~~ \text{and} ~~\mathcal{C}^{\prime}$ versus $M$}.
\label{Specificheat}
\end{figure}
We plot in Fig. (\ref{Specificheat}) the specific heat versus the BH mass (putting $n=4$ an an example) and found that the specific heat of rainbow BH
diverges at a point at which the BH temperature reaches its maximum value and then it decreases to zero  when
the mass of the BH reaches its minimal value (i.e \emph{remnant}). At this point, specific heat vanishes and hence
the BH do not exchange heat with the surrounding space leaving a remnant.

\begin{figure}
\includegraphics[scale=0.35,angle=0]{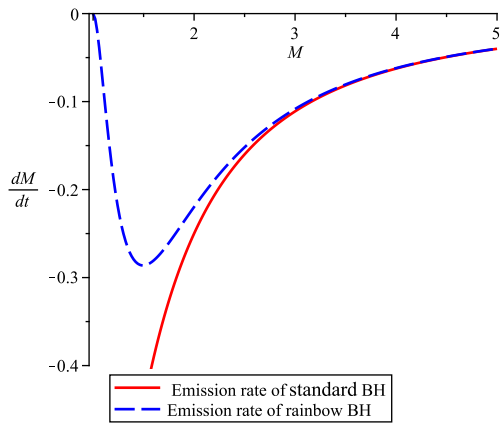}
\caption{$\frac{dM}{dt}_{\text{standard}}~~ \text{and} ~~\frac{dM}{dt}_{\text{rainbow}}$ versus $M$}.
\label{dMdt}
\end{figure}

Now, we use the rainbow Hawking temperature to calculate the emission rate of the rainbow BH
using Stefan-Botlzmann law considering the energy loss was dominated by
photons. This is found to be
\bea
\frac{dM}{dt}_{(\text{rainbow})}=\frac{dM}{dt}_{(\text{standard})} \le(1-\eta \frac{E_p^n}{(2M)^n }\ri)^2, \la{rate2}
\eea
This means that the emission rate of the rainbow BH vanishes when the BH reaches its minimal value
$M=M_{min}=\frac{1}{2}\eta^{1/n} E_p$. We show in Fig. (\ref{dMdt}) (putting $n=4$) how the picture of the emission rate of the BH
got changed in gravity rainbow. In the standard picture, the emission rate goes to infinity as the mass of
the BH tends to zero. In rainbow gravity, the emission rate of rainbow BH do not diverge at all, and it just go to zero
when the rainbow BH reaches its minimum value which can be called as a BH remnant.
Since remnant of  black holes need not possess the horizon, we think that our result may
ameliorate information loss problem\cite{Biswas:2011ar}

It is worth comparing our results with previous analysis of black hole evaporation with MDR and GUP that has have
been investigated in \cite{AmelinoCamelia:2005ik}. It has been found in \cite{AmelinoCamelia:2005ik}
that the modified temperature of evaporating black hole due to MDR diverges when the mass of
the black hole decreases to a Planck-scale value instead of having divergence only when the mass tends to zero in the standard case.
It is clear this picture is different from the picture that we obtained in our work in Eq. (\ref{modT}) in which the
temperature do not suffer any divergence as the mass of the black hole approaches Planck scale. On the contrary, the Hawking temperature
for rainbow BH in Eq. (\ref{modT}) increases as the mass of the black hole decreases until it reaches maximum value, and then decreasing
to be zero when the black hole reaches a remnant case at which the black hole cannot exchange heat with the
surrounding space.

\section{Conclusions}
\label{conclusions}

In this work we studied the thermodynamical picture of rainbow black hole. We
found that gravity rainbow may lead to a new mass-temperature relation and define a minimum mass
and maximum temperature for rainbow black hole predicting the existence of \emph{black hole remnant}.
We found that the end-point of Hawking radiation is not a catastrophic because rainbow functions
that are proposed by Amelino-Camelia, et al. in \cite{amerev, AmelinoCamelia:1996pj}
imply the existence of BH remnants at which the specific heat
vanishes and, therefore, the BH cannot exchange heat with the
surrounding space. The gravity rainbow  prevents BHs from evaporating
completely, just like the standard uncertainty principle
prevents the hydrogen atom from collapsing\cite{Adler,Cavaglia:2003qk}.
Our result agree with results obtained in the framework of generalized uncertainty principle
and black hole physics which also predicted the existence of black hole remnants\cite{Adler,Cavaglia:2003qk}.
Since remnant of  black holes does not possess the horizon, we think that our result may ameliorate information loss problem\cite{Biswas:2011ar}. 
In the future, it would be appropriate to generalize the calculations in extra
dimensions to investigate the possibilities to see the remnants of rainbow black holes at LHC.


\subsection*{Acknowledgments}

The author gratefully thanks the anonymous referee for enlightening comments and suggestions
which substantially helped in proving the quality of the paper.
The author gratefully thanks Anupam Mazumdar for enlightening
comments on the manuscript. The research of the author is supported by 
Benha University (www.bu.edu.eg) and CFP in Zewail City, Egypt.




\begin{thebibliography}{100}

\bibitem{amerev}
  G.~Amelino-Camelia,
  Living Rev.\ Rel.\  {\bf 16}, 5 (2013)
  [arXiv:0806.0339 [gr-qc]].




\bibitem{AmelinoCamelia:1996pj}
  G.~Amelino-Camelia, J.~R.~Ellis, N.~E.~Mavromatos and D.~V.~Nanopoulos,
  Int.\ J.\ Mod.\ Phys.\ A {\bf 12}, 607 (1997)
  [hep-th/9605211].

%
%
%

\bibitem{'tHooft:1996uc}
  G.~'t Hooft,
  Class.\ Quant.\ Grav.\  {\bf 13}, 1023 (1996)
  [gr-qc/9601014].

\bibitem{LIstring}
V. A. Kostelecky and S. Samuel, Phys. Rev. D {\bf 39}, 683 (1989).

\bibitem{Gambini:1998it}
  R.~Gambini and J.~Pullin,
  Phys.\ Rev.\ D {\bf 59}, 124021 (1999)
  [gr-qc/9809038].

\bibitem{Carroll:2001ws}
  S.~M.~Carroll, J.~A.~Harvey, V.~A.~Kostelecky, C.~D.~Lane and T.~Okamoto,
  Phys.\ Rev.\ Lett.\  {\bf 87}, 141601 (2001)
  [hep-th/0105082].

\bibitem{AmelinoCamelia:1997gz}
  G.~Amelino-Camelia, J.~R.~Ellis, N.~E.~Mavromatos, D.~V.~Nanopoulos and S.~Sarkar,
  Nature {\bf 393}, 763 (1998)
  [astro-ph/9712103].

\bibitem{AmelinoCamelia:1997jx}
  G.~Amelino-Camelia, J.~Lukierski and A.~Nowicki,
  Phys.\ Atom.\ Nucl.\  {\bf 61}, 1811 (1998)
  [Yad.\ Fiz.\  {\bf 61}, 1925 (1998)]
  [hep-th/9706031].

\bibitem{AmelinoCamelia:1999wk}
  G.~Amelino-Camelia, J.~Lukierski and A.~Nowicki,
  Int.\ J.\ Mod.\ Phys.\ A {\bf 14}, 4575 (1999)
  [gr-qc/9903066].
  
\bibitem{AmelinoCamelia:2002dx}
  G.~Amelino-Camelia,
  New J.\ Phys.\  {\bf 6}, 188 (2004)
  [gr-qc/0212002].

\bibitem{AmelinoCamelia:2000mn}
  G.~Amelino-Camelia,
  Int.\ J.\ Mod.\ Phys.\ D {\bf 11}, 35 (2002)
  [gr-qc/0012051]; J.~Magueijo and L.~Smolin,
  Phys.\ Rev.\ D {\bf 67}, 044017 (2003)
  [gr-qc/0207085].

\bibitem{Magueijo:2002xx}
  J.~Magueijo and L.~Smolin,
  Class.\ Quant.\ Grav.\  {\bf 21}, 1725 (2004)
  [gr-qc/0305055].

\bibitem{Galan:2004st}
  P.~Galan and G.~A.~Mena Marugan,
  Phys.\ Rev.\ D {\bf 70}, 124003 (2004)
  [gr-qc/0411089];
  J.~Hackett,
  Class.\ Quant.\ Grav.\  {\bf 23}, 3833 (2006)
  [gr-qc/0509103]; F.~Girelli, S.~Liberati and L.~Sindoni,
  Phys.\ Rev.\ D {\bf 75}, 064015 (2007)
  [gr-qc/0611024]; C.~-Z.~Liu and J.~-Y.~Zhu,
  Gen.\ Rel.\ Grav.\  {\bf 40}, 1899 (2008)
  [gr-qc/0703055 [GR-QC]];
   H.~Li, Y.~Ling and X.~Han,
  Class.\ Quant.\ Grav.\  {\bf 26}, 065004 (2009)
  [arXiv:0809.4819 [gr-qc]]; R.~Garattini and G.~Mandanici,
  Phys.\ Rev.\ D {\bf 85}, 023507 (2012)
  [arXiv:1109.6563 [gr-qc]]; R. Garattini and F. S.N. Lobo, Phys.\ Rev.\ D {\bf 85}, 024043 (2012);
  R. Garattini and G. Mandanici, Phys. Rev. D {\bf 83}, 084021 (2011);
   J.~-J.~Peng and S.~-Q.~Wu,
  Gen.\ Rel.\ Grav.\  {\bf 40}, 2619 (2008)
  [arXiv:0709.0167 [hep-th]].

\bibitem{FRWRainbow}
 Y.~Ling,
  JCAP {\bf 0708}, 017 (2007)
  [gr-qc/0609129];  Y.~Ling and Q.~Wu,
  Phys.\ Lett.\ B {\bf 687}, 103 (2010)
  [arXiv:0811.2615 [gr-qc]].

\bibitem{Barrow:2013gia}
  J.~D.~Barrow and J.~Magueijo,
  arXiv:1310.2072 [astro-ph.CO];   G.~Amelino-Camelia, M.~Arzano, G.~Gubitosi and J.~Magueijo,
  Phys.\ Rev.\ D {\bf 88}, 041303
  [arXiv:1307.0745 [gr-qc]].

\bibitem{Awad:2013nxa}
  A.~Awad, A.~F.~Ali and B.~Majumder,
  JCAP {\bf 1310}, 052 (2013)
  [arXiv:1308.4343 [gr-qc]].

\bibitem{Hawking:1974sw}
  S.~W.~Hawking,
  Commun.\ Math.\ Phys.\  {\bf 43}, 199 (1975)
  [Erratum-ibid.\  {\bf 46}, 206 (1976)];  S.~W.~Hawking,
  Phys.\ Rev.\ D {\bf 13}, 191 (1976); J.~D.~Bekenstein,
  Lett.\ Nuovo Cim.\  {\bf 4}, 737 (1972).
\bibitem{GRbooks}
Oyvind Gron, Einstein's general theory of relativity: with modern applications in cosmology. Springer, 2007;
R.M.Wald, General relativity (University of Chicago
Press; Chicago, 1984).

\bibitem{Bekenstein:1973ur}
  J.~D.~Bekenstein,
  Phys.\ Rev.\ D {\bf 7}, 2333 (1973).

\bibitem{Adler}
 R.~J.~Adler, P.~Chen, D.~I.~Santiago,
  Gen.\ Rel.\ Grav.\  {\bf 33}, 2101-2108 (2001).
  [gr-qc/0106080].

\bibitem{Cavaglia:2003qk}
  M.~Cavaglia, S.~Das, R.~Maartens,
  Class.\ Quant.\ Grav.\  {\bf 20}, L205-L212 (2003).
  [hep-ph/0305223];
  M.~Cavaglia, S.~Das,
  Class.\ Quant.\ Grav.\  {\bf 21}, 4511-4522 (2004).
  [hep-th/0404050].

\bibitem{Niemeyer:2001xk}
  J.~C.~Niemeyer,
  Phys.\ Rev.\ D {\bf 65}, 083505 (2002)
  [astro-ph/0111479]; A. Kempf, J.Phys. {\bf A 30} (1997) 2093 [arXiv:hep-th/9604045].



\bibitem{Eliasentropy0}
A.~J.~M.~Medved, E.~C.~Vagenas,
  Phys.\ Rev.\  D {\bf 70}, 124021 (2004).
  [hep-th/0411022].

\bibitem{Eliasentropy1}
B.~Majumder,
  Phys.\ Lett.\ B {\bf 703}, 402 (2011).
[arXiv:1106.0715 [gr-qc]].

\bibitem{AmelinoCamelia:2004xx0}
  G.~Amelino-Camelia, M.~Arzano and A.~Procaccini,
  Phys.\ Rev.\ D {\bf 70}, 107501 (2004)
  [gr-qc/0405084];
E.M. Lifshitz, L.P. Pitaevskii and V.B. Berestetskii, Landau-Lifshitz Course of Theoretical
Physics, Volume 4: Quantum Electrodynamics, (Reed Educational and Professional
Publishing, 1982).

\bibitem{AmelinoCamelia:2005ik}
  G.~Amelino-Camelia, M.~Arzano, Y.~Ling and G.~Mandanici,
  Class.\ Quant.\ Grav.\  {\bf 23}, 2585 (2006)
  [gr-qc/0506110].
  
\bibitem{Biswas:2011ar}
  T.~Biswas, E.~Gerwick, T.~Koivisto and A.~Mazumdar,
  Phys.\ Rev.\ Lett.\  {\bf 108}, 031101 (2012)
  [arXiv:1110.5249 [gr-qc]].







\end{thebibliography}
\end{document}